\documentclass[11pt,twoside]{article}
\usepackage{asp2010}

\resetcounters

\begin{document}

\title{Uncovering the pulsating photospheres of Mira stars through near-IR interferometry: a case study on R Vir}
\author{M. Hillen$^1$, T. Verhoelst$^1$, B. Acke$^1$ and P. Degroote$^1$
\affil{$^1$Instituut voor Sterrenkunde, KU Leuven, Celestijnenlaan 200D, 3001 Leuven, Belgium}}

\begin{abstract}
We present the methodology and some preliminary results of our study of the relationship between a Mira's pulsating photosphere and its surrounding molecular layer(s) throughout several pulsation cycles, based on spatially resolved data. Our dataset consists of archival narrow-band observations in the near-infrared H and K bands obtained with the Palomar Testbed Interferometer between 1999 and 2006, extended with a few nights of VLTI AMBER low spectral resolution data and near-infrared SAAO photometry. The fitted model is the geometric star + layer model proposed by Perrin et al. (2004), in which the physical parameters (diameter and temperature of star and layer; wavelength dependent optical depth of the layer) are given a sinusoidal time dependence.
\end{abstract}

\section{Introduction}
The issue of Mira pulsation modes has long been an issue of diameters. Velocity amplitude measurements \citep{Bessell1996} and period-luminosity relations \citep{Wood2000} are evidence in favor of the fundamental pulsation mode hypothesis, while measured diameters \citep{Tuthill1994} in the past pointed towards a first overtone pulsation. Moreover, the scales of the varying Mira diameters with wavelength can not be accounted for by the trend in continuum opacity reaching a minimum around 1.6~$\mu m$. This changed when \citet{Mennesson2002} \citep[and later][]{Perrin2004} found that the measured diameters are actually biased by the presence of a molecular layer (consisting mainly of CO and/or H$_2$O) at up to a few stellar radii above the photosphere. The observed strong diameter variations with wavelength are in this simple geometrical model explained by the varying transparency of the shell and the wavelength dependence of black body radiation. \citet{LeBouquin2009} later confirmed this basic picture by performing a model-independent image reconstruction on an interferometric dataset continuous in both the spectral (1.5 - 2.4~$\mu m$) and spatial domain for the Mira variable T Lep. 

Although the general picture is now well established, a lot of questions related to the molecular layer remain unresolved. What is its role in driving the stellar wind? What are its exact sources of opacity? \citet{LeBouquin2009} needed a source of continuum opacity ($\tau = 0.2$) on top of CO and H$_2$O to explain the optical depths required by the fit to the interferometric observables. How do the photosphere and molecular layer vary with the pulsation? How does this time dependence compare with predictions of pulsation theory? Are Miras really pulsating in the fundamental mode? 

The immediate goal of our study is to quantify the temporal behavior of photosphere and molecular layer for a sample of Mira variables. To this end, we extend the analytical model of \citet{Perrin2004} by assuming a sinusoidal time dependence for the physical parameters $\theta_{\star}$, $T_{\star}$, $\theta_{layer}$, $T_{layer}$ and $\tau(\lambda)$. 

\section{Data and finding the best fit}
Due to its large and diverse sample ($\sim 100$) of observed Mira variables, and excellent temporal coverage (about 8~yrs), we tried to apply our model to the publicly released Palomar Testbed Interferometer (PTI) dataset. The PTI is a fixed three telescope interferometer (baselines of 86m, 87m, and 110m roughly oriented NW, SW, and NS respectively) operating in five narrow spectral bands covering 2.0 - 2.4~$\mu m$ or (i.e. not within the same night) four channels covering 1.50 - 1.75~$\mu m$. 

R~Vir is the second best observed Mira in the PTI archive and has about 275 squared visibilities per wavelength, spread over 12 pulsation cycles. \citet{Eisner2007} previously applied the same model to a spectrally dispersed interferometric Keck dataset of R Vir during rise to maximum visual light. The Keck data, which have a higher spectral resolution ($R \sim 230$ vs. $R \sim 30$) but carry less spatial information, are consistent with the PTI data at similar phases. We extended our dataset with three VLTI-AMBER points in low spectral resolution on triplet A0-G1-K0 during GTO time in February 2009. In contrast to what was presented on the conference, there is no discrepancy between these AMBER data and PTI data at similar phases (the presented discrepancy was caused by a local software problem in the calibration process).

To fit the data we make use of two independent methods: a standard Nelder-Mead Simplex algorithm and a genetic algorithm. Our main concern at the moment are the degeneracies the 2-component model gives: very similar visibility curves can be produced with very different values for the physical parameters. From our fitting and simulations we conclude that the PTI dataset has insufficient spatial information at high enough accuracy to constrain the model. The AMBER data do constrain the model, but only at that particular phase. Hopefully, combining the two datasets gives enough constraints to disentangle the temporal behavior of the different components in the model.

\acknowledgements The Palomar Testbed Interferometer is operated by the NASA Exoplanet Science Institute on and the PTI collaboration and was constructed with funds from the Jet Propoulsion Laboratory, Caltech as provided by the National Aeronautics and Space Administration. 

\bibliography{Hillen}

\end{document}